# Magnetism of Edge Modified Nano Graphene

Norio Ota, Narjes Gorjizadeh* and Yoshiyuki Kawazoe*

*Pure and Applied Sciences, University of Tsukuba, 1-1-1 Tenoudai, Tsukuba 305-8573 JAPAN*
*\* Institute for Materials Research, Tohoku University, 2-1-1 Katahira, Aoba-ku, Sendai 980-8577 JAPAN*

In order to study a magnetic principle of carbon based materials, multiple spin state of zigzag edge modified graphene molecules are analyzed by the first principle density functional theory to select suitable modification element. Radical carbon modified $C_{64}H_{17}$ shows that the highest spin state is most stable, which arises from two up-spin's tetrahedral molecular orbital configuration at zigzag edge. In contrast, oxygen modified $C_{59}O_5H_{17}$ show the lowest spin state to be most stable due to four spins cancellation at oxygen site. Boron modified $C_{59}B_5H_{22}$ have no $\pi$-molecular orbit at boron site to bring stable molecular spin state to be the lowest one. Whereas, $C_{59}N_5H_2$ have two $\pi$-electrons, where spins cancel each other to give the stable lowest spin state. Silicon modified $C_{59}Si_5H_{27}$ and Phosphorus modified $C_{59}P_5H_{22}$ show curved molecular geometry due to a large atom insertion at zigzag site, which also bring complex spin distribution. Radical carbon and dihydrogenated carbon modification are promising candidates for designing carbon-based magnetic materials.

## 1. Introduction

Magnetism of graphene and graphite-like materials are very attractive both in scientific aspect of $\pi$-electron oriented magnetism and both in technological aspect of light weight ecological magnet[1)-4)] and novel spintronic devices[5)-9)]. These years, several experiments[10)-15)] have suggested a capability of room-temperature ferromagnetism. Recently, H. Ohldag et al[15)] opened that proton ion implanted graphite shows a strong surface magnetism with a saturation magnetization of 12-15emu/g at 300K. Also T. Saito et al[12)] predicted ferromagnetic graphite-like carbon material by a pyrolysis method. Those experiments encourage us to create new attractive materials. However, basic origin of magnetism and advanced mechanism of bulk ferro-magnetism are not clear yet. This paper focuses on a former issue, that is, to build a model of magnetic origin in nano-meter scaled graphene molecules with chemically modified zigzag edges, and to find out which modification has a potential for strong magnetism. A first principle density functional theory (DFT) is applied for detailed analysis.

There were many theoretical predictions on graphene zigzag edge carbon magnetism due to localized density states near Fermi energy[16)-20)]. Focusing on infinite length graphene ribbon, Fujita et al[16)] applied tight binding model resulting antiferromagnetic feature with total magnetization zero. Rectangular shaped graphene has been extensively studied[21)-24)]. The singlet state (zero magnetization) arises by both side zigzag edge modifications with various species.[23)] Whereas, Kusakabe and Maruyama[25)26)] proposed an asymmetric infinite length ribbon model showing ferrimagnetic behavior with non-zero total magnetization.

Our previous papers reported multiple spin state analysis in asymmetric graphene molecule, especially, in a dihydrogenated zigzag edge graphene molecule[27)28)]. These models have resulted that in every molecule the highest spin state is the most stable molecular energy. This paper enhances modified elements following a periodic atomic table as like boron, radical carbon, nitrogen, oxygen, silicon, and phosphorus. We like to predict which one is favorable for realizing strong magnetism.

## 2, Model Molecules

Origin of magnetism of conventional ferromagnetic materials like Fe, Co, Ni etc. is well known to be brought by 3d-atomic orbital configuration. For example, as simply imaged in Fig.1, five 3d-electrons of $Fe^{3+}$ show five parallel up-spins by Hund's rule[31)32)]. Bulk ferromagnetism needs more deep and complex discussion. Whereas in case of carbon based material, studies both on basic origin of magnetism and on bulk ferromagnetism are not clear yet. This paper focuses the former part, that is, a model of magnetic origin. Our idea is to start from an asymmetric graphene molecule. As an analogy of Fe3+, our model molecule has chemically modified five zigzag edges. Periodic table like modification is done by BH (chemical formula: $C_{59}B_5H_{22}$), radical carbon C: ($C_{64}H_{17}$), dihydrogenated carbon $CH_2$ ($C_{64}H_{27}$), NH ($C_{59}N_5H_{22}$), Oxygen ($C_{59}O_5H_{17}$), $SiH_2$ ($C_{59}Si_5H_{27}$), and PH ($C_{59}P_5H_{22}$). For comparing magnetism, common condition is asked to keep five unpaired electrons inside of every molecule. These five electrons give three possible total molecular spin state Sz (perpendicular component to a molecular plane) of 1/2, 3/2, or 5/2 as shown in Fig.1. Question is which spin state is favorable for stabilizing every molecule, that is, which spin state gives the lowest total molecular energy. Of course, molecule with stable Sz=5/2 may give a possible building block for materials with strong magnetism. We like to find out favorable modification elements and study basic origin for designing future bulk ferromagnetic material.

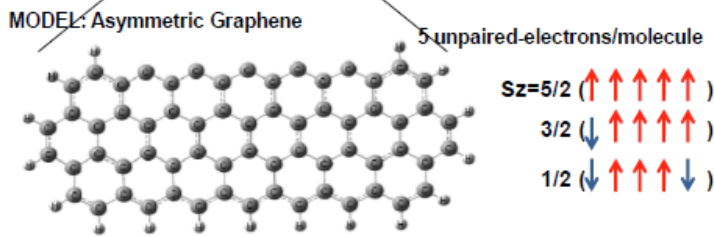

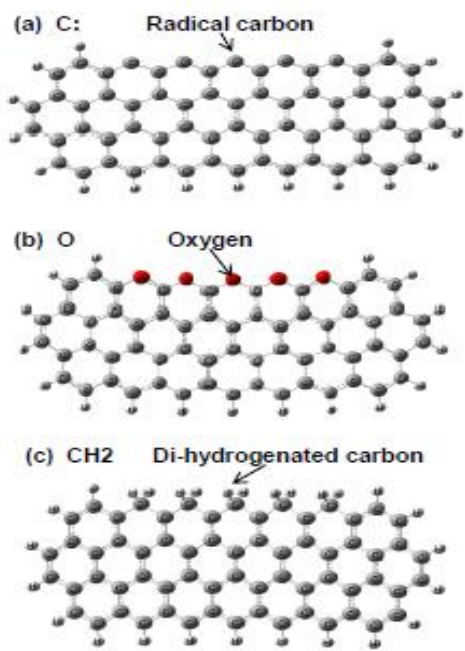

Fig.1 Asymmetric graphene molecule is one candidate modeling a magnetic origin of carbon system.

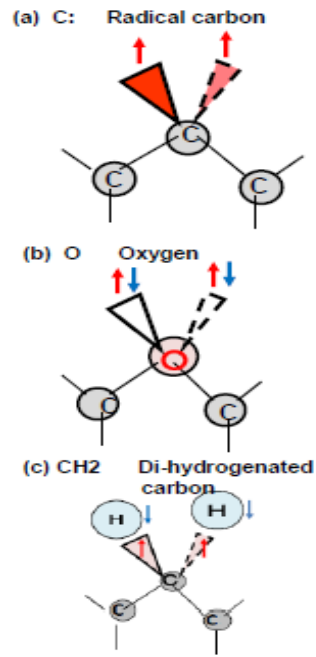

Fig.3 Tetrahedral configuration of electrons around zigzag edge site in three cases

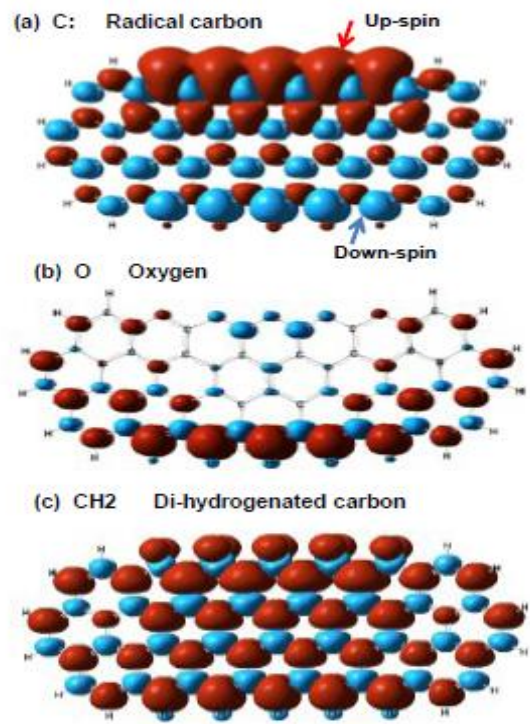

Fig.2 Typical asymmetric graphene molecules: (a) $C_{64}H_{17}$ with five radical carbon edges (b) $C_{59}O_5H_{17}$ with five oxygen edges (c) $C_{64}H_{27}$ with five dihydrogenated carbon edges.

Fig.4 Spin density contour surface of (a) $C_{64}H_{17}$ with Sz=5/2, (b) $C_{59}O_5H_{17}$ with Sz=1/2, and (c) $C_{64}H_{27}$ with Sz=5/2. These show stable spin state in every molecule.

## 3. Calculation Methods

In order to clarify magnetism, we have to obtain (i) spin density map, (ii) total molecular energy and (iii) optimized atom arrangement depending on respective given spin state Sz. Density functional theory (DFT) [31)32)] based generalized gradient approximation method (GGA-UPBEPBE)[33)] is applied for those calculations. Atomic orbital basis is 6-31G [34)]. For obtaining accurate self consistent field energy, requested convergence on root mean square density matrix was less than 10E-8 within 128 cycles.

## 4, Radical-carbon, Oxygen, and Dihydrogenated-carbon Modified Molecules

Radical carbon edge $C_{64}H_{17}$ molecule in Fig.2 (a) is an example of typical asymmetric graphene. DFT calculation reveals that the highest spin state Sz=5/2 is the lowest molecular energy state, that is, the most stable one as shown in Fig.5. Next one is Sz=3/2, and unstable solution is Sz=1/2, which have up-up and down-down complex spin pairs inside of a molecule. Those adjoined parallel spin pairs give unnecessary exchange energy and finally increase total molecular energy as resulted in Fig.5. Also, it should be noted that there appear twice a large up-spin cloud (red in color, dark gray) at zigzag edge radical carbon site as illustrated in Fig.4 (a). All figures of spin density contour surface is drawn at 0.001e/A³. Based on such remarkable spin configuration, we assumed tetrahedral molecular orbits as imaged in Fig.3 (a). Two unpaired electrons occupy two identical tetrahedral orbits. By a molecular Hund's rule[29)30)], they both become up-spin.

Based on such concept, we can estimate the other advanced case with four electrons in the same tetrahedral molecular orbits as illustrated in Fig.3 (b). Capable atom with four electrons at zigzag edge site is oxygen as $C_{59}O_5H_{17}$ in Fig.2 (b). Those four electrons should occupy two tetrahedral orbits as two sets of up-down spin pairs by Pauli principle as shown in Fig.3 (b). Spin cloud may become so weak or zero around zigzag edge site, and may bring weak magnetism to a molecule itself. This estimation is supported by DFT calculation. Result is the most stable spin state to be Sz=1/2 (Fig.5), and spin cloud around zigzag edge is not so dense as illustrated in Fig.4 (b).

What happens when two hydrogen bond with tetrahedral molecular orbits. Two electrons from one carbon and two electrons from two hydrogen may create new molecular orbits as shown in Fig.3(c). Simple expectation is a cancelation of spin as two sets of up-down pair. However, calculated result of dihydrogenated carbon molecule $C_{64}H_{27}$ (see Fig.2 (c)) is an unexpected one, that is, there appear up-spin cloud at attached two hydrogen site (see Fig.4 (c)), and the most stable spin state is Sz=5/2 (Fig.5). This is the general result even if changing numbers of zigzag edges[27)28)]. We assumed that electron cloud of hydrogen is just sphere and small, which bring remained unpaired up-spin cloud at hydrogen (two sites) and down-spin cloud at zigzag edge carbon site as shown in Fig.4 (c).

## 5, Boron-hydrogenated and Nitrogen-hydrogenated Molecules

Polycyclic aromatic hydrocarbon (PAH) makes $\pi$-electron network. Typical example is $C_{64}H_{22}$, which has no unpaired electrons resulting total molecular spin polarization Sz=0 and shows diamagnetic nature as shown in Fig.6(a) and Fig.8(a).

When those five zigzag edge carbons are substituted by boron as like $C_{59}B_5H_{22}$, there disappear five $\pi$-electrons. Whereas, in view of a whole molecule, such lack of five $\pi$-electrons brings five unpaired electrons to be an origin of molecular magnetism. We should discuss multiple spin state stability. DFT calculation predicted that the most stable spin state is Sz=1/2 (Fig.5) and almost no spin cloud around boron (Fig.8(b)). Such result is simply explained by a disappearance of $\pi$-electron at boron site (see Fig.7 (b)). There are no spin source around boron, and inside of a molecule up-down spin pair is favorable for reducing total energy.

Whereas in case of nitrogen modified $C_{59}N_5H_{22}$, as imaged in Fig.7 (c), zigzag edge site $\pi$-orbit will be occupied by two electrons. By a Pauli principle, those two electrons play up-down spin pair as shown in Fig.7(c). This brings spin-cancelled zigzag edge site, and may give weak molecular magnetism. Such speculation is supported by detailed DFT calculation. Result is clear that the most stable spin state is the lowest spin state to be Sz=1/2 (Fig.5), and the spin density map (Fig.8(c)) is very similar with boron modified case.

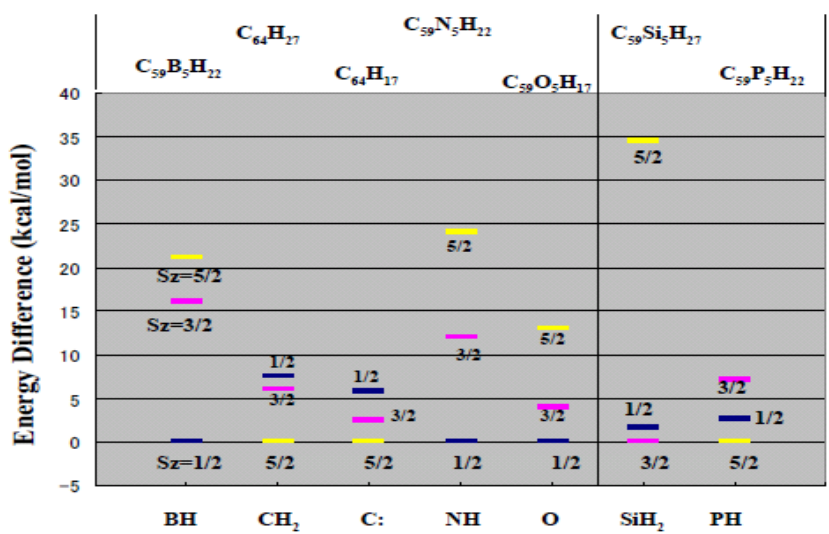
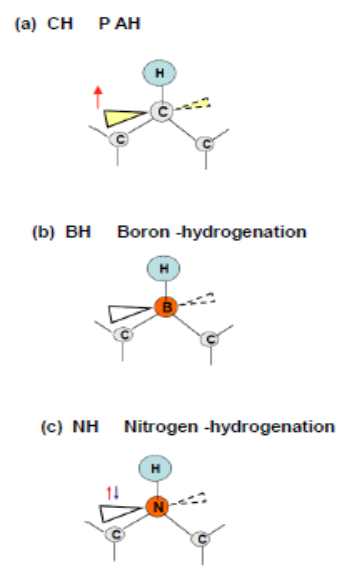

**Fig.5** Total molecular energy difference between spin states in every molecule. Energy of the most stable spin state is scaled to be zero.

**Fig.7** Hydrogenated molecule's electrons configuration at zigzag edge site in (a) $C_{64}H_{22}$ with one $\pi$-electron, (b) $C_{59}B_5H_{22}$ with no $\pi$-electron, and (c) $C_{59}N_5H_{22}$ with two $\pi$-electrons.

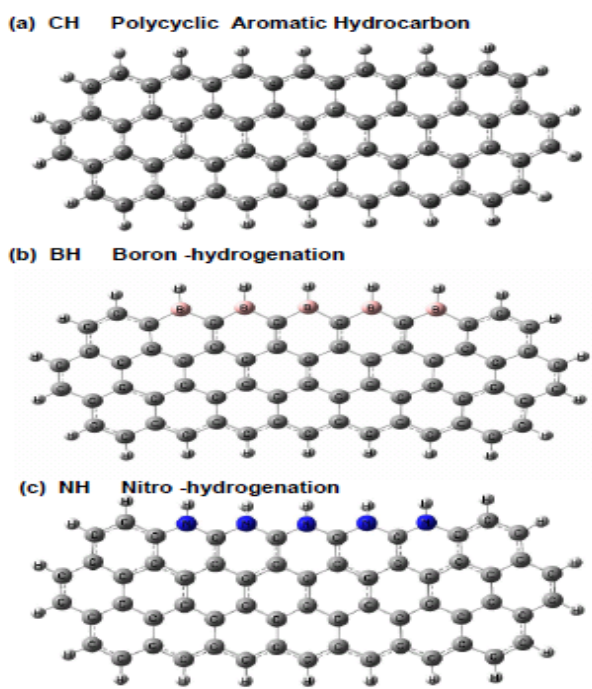

**Fig.6** Hydrogenated zigzag edge molecules:
(a) polycyclic aromatic hydrocarbon (PAH) $C_{64}H_{22}$,
(b) boron substituted $C_{59}B_5H_{22}$, and (c) nitrogen substituted $C_{59}N_5H_{22}$.

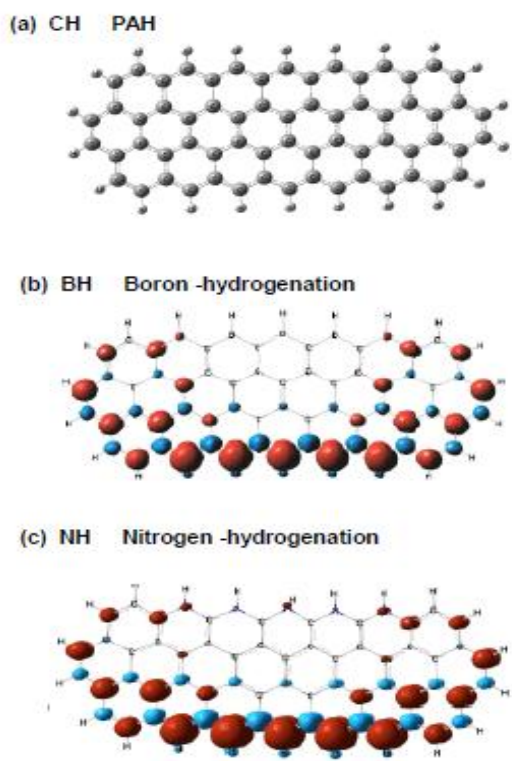

**Fig.8** Spin density contour surface of (a) $C_{64}H_{22}$ with Sz=0, (b) $C_{59}B_5H_{22}$ with Sz=1/2, and $C_{59}N_5H_{22}$ with Sz=1/2, which are the most stable spin state in every molecule.

## 6, Silicon-dihydrogenated and Phosphorus- hydrogenated Molecules

Silicon-dihydrogenated $C_{59}Si_5H_{27}$ is an analogy of carbon-dihydrogenated $C_{64}H_{27}$. Simple expectation will be similar result. However, we obtained somewhat complicated one. Remarkable difference is an atomic arrangement as shown in Fig.9 (a) and (b). There appears a curved geometry. This reason is simple that an atomic radius of silicon is larger than carbon, which causes a strong curving force to bend a total molecular structure. Stable spin state is decided by a total balance of atomic configuration and spin-spin interaction, which finally gives complex spin density map.

The most stable spin state is Sz=3/2, whereas unstable one is 5/2 as shown in Fig.5. Spin density map is illustrated in Fig.10(d), where tetrahedral spin configuration is observed at silicon site as simply imaged in Fig.10 (c).

Phosphorus-hydrogenated $C_{59}P_5H_{22}$ has also curved geometry as shown in Fig.11 (a) and (b). Such complex geometry affects multiple spin state stability. The highest spin state Sz=5/2 is the most stable one (see Fig.5). Spin cloud around phosphorus shows tetrahedral configuration illustrated in Fig.11(c), which is different with a simple analogy with nitrogen-hydrogenation case.

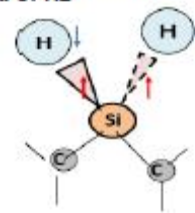

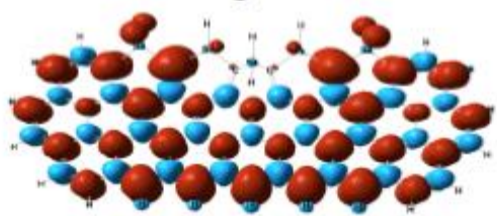

**Fig.10** Results of $C_{59}Si_5H_{22}$ with Sz=3/2: (c) tetrahedral molecular orbital image at zigzag edge, and (d) spin density contour surface.

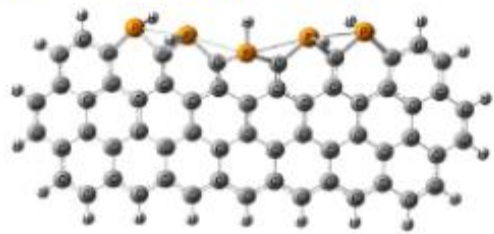

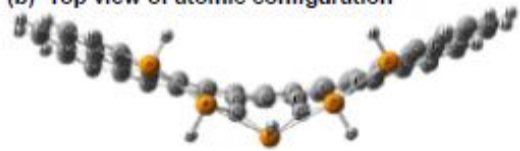

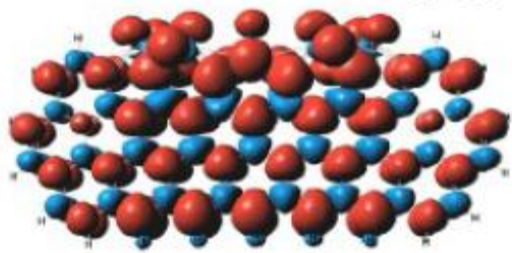

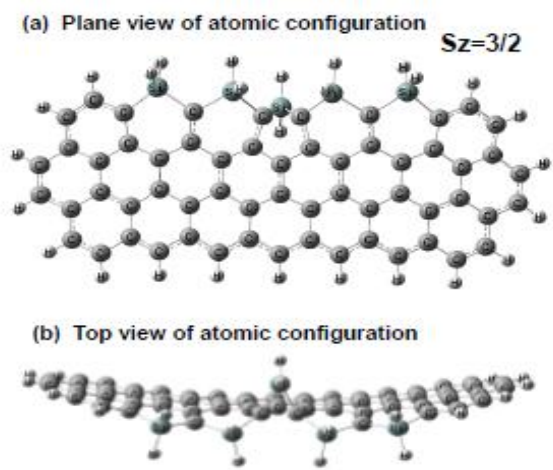

**Fig.9** Results of $C_{59}Si_5H_{22}$ with Sz=3/2: (a) atomic configuration plane view ,(b) and top view.

**Fig.11** Results of $C_{59}P_5H_{22}$ with Sz=5/2: (a) atomic configuration plane view, (b) top view, and (c) spin contour surface.

# 7, Conclusion

To study magnetic origin of graphene is very important for designing future bulk carbon-based magnetic materials. We modeled zigzag edge modified graphene molecules and analyzed multiple spin state to select suitable modification element by the first principle density functional theory DFT. Results are,

(1) Radical carbon modified $C_{64}H_{17}$ and carbon-dihydrogenated $C_{64}H_{27}$ show strong magnetism. The highest spin state is the most stable, which arises from tetrahedral molecular orbits at zigzag edge site. Spin configuration is governed by Hund's rule.

(2) Oxygen modified $C_{59}O_5H_{17}$ has four electrons, which occupy tetrahedral two orbits as two sets of up-down spin pairs. Canceled spin brings weak magnetism. The lowest spin state is the most stable one.

(3) Boron modified $C_{59}B_5H_{22}$ have $\pi$-molecular orbit configuration at boron, but actually no $\pi$-electron cloud at zigzag edge site. Whereas nitrogen modified $C_{59}N_5H_{22}$ have two $\pi$-electrons, which cancels each other as up-down spin pair by Pauli principle. In both cases, there appear almost no spin polarized cloud at zigzag edge, which brings weak magnetism. The most stable spin state is the lowest one.

(4) Both silicon modified $C_{59}Si_5H_{27}$ and phosphorus modified $C_{59}P_5H_{22}$ show curved geometry by a large Si or P atom insertion to zigzag edge site. Such complex geometry affects to give the stable spin state, that is, Sz=3/2 in $C_{59}Si_5H_{27}$, whereas Sz=5/2 in $C_{59}P_5H_{22}$.

It is concluded that favorable candidates are radical carbon and dihydrogenated carbon modification.